\documentclass[journal]{IEEEtran}
\usepackage{jrn_setup}
\bibliography{EventTiming} % if biblatex

\AddToHook{shipout/background}{%
  \put(0.5in,-0.5in){%
    \parbox[c]{\textwidth}{\centering \normalfont\Large\bfseries DOI \href{https://doi.org/10.1109/JBHI.2024.3373124}{10.1109/JBHI.2024.3373124}. \copyright 2024 IEEE. \\ \normalsize This article has been accepted for publication in IEEE Journal of Biomedical and Health Informatics}
  }
}  

\begin{document}
\IEEEpubid{\copyright 2024 IEEE}

\title{Cardiac Valve Event Timing in Echocardiography \\[0.15cm] using Deep Learning and Triplane Recordings}

\author{Benjamin~Strandli~Fermann,~\IEEEmembership{Student Member,~IEEE,}
        John~Nyberg, 
        Espen~W.~Remme,
        Jahn~Frederik~Grue,
        Helén~Grue,
        Roger~Håland,
        Lasse~Lovstakken,
        H{\aa}vard~Dalen,
        Bj{\o}rnar~Grenne,
        Svein~Arne~Aase,
        Sten~Roar~Snare,
        and~Andreas~{\O}stvik,~\IEEEmembership{Member,~IEEE}% <-this % stops a space
\thanks{Manuscript received 3 July, 2023; revised 2 February 2024, 2023; accepted  29 February 2024. This work was funded by the Research Council of Norway (Project number 237887 and 312542, and PROCARDIO), Mid-Norway regional health authority and GE Vingmed Ultrasound.}
\thanks{B. S. Fermann, S. A. Aase, S. R. Snare are with GE Vingmed Ultrasound, 3183 Horten, Norway.
A.~{\O}stvik, B. Grenne, H. Dalen, J. F. Grue, J. Nyberg and L. Lovstakken are with the Dept. of Circulation and Medical Imaging at the Norwegian University of Science and Technology, 7030 Trondheim, Norway.
H. Grue and R. H{\aa}land are with the Dept. of Cardiology, Oslo University Hospital, Rikshospitalet, 0372 Oslo, Norway.
E. W. Remme is with the Intervention Centre, Oslo University Hospital, Rikshospitalet, 0372 Oslo, Norway, and also with the Institute for Surgical Research, Oslo University Hospital, Rikshospitalet, 0372 Oslo, Norway.
B. S. Fermann is also with the Dept. of Informatics at University of Oslo, 0316 Oslo, Norway.
A.~{\O}stvik is also with SINTEF Digital, Dept. of Health Research, 7465 Trondheim, Norway.
B. Grenne and H. Dalen are also with Clinic of Cardiology, St. Olavs hospital, 7030 Trondheim, Norway.}%
}

% \markboth{Journal of Biomedical and Health Informatics}%
% {} % For odd pages when two-sided

\maketitle

\begin{abstract}
Cardiac valve event timing plays a crucial role when conducting clinical measurements using echocardiography. However, established automated approaches are limited by the need of external electrocardiogram sensors, and manual measurements often rely on timing from different cardiac cycles. Recent methods have applied deep learning to cardiac timing, but they have mainly been restricted to only detecting two key time points, namely end-diastole (ED) and end-systole (ES). In this work, we propose a deep learning approach that leverages triplane recordings to enhance detection of valve events in echocardiography. Our method demonstrates improved performance detecting six different events, including valve events conventionally associated with ED and ES. Of all events, we achieve an average absolute frame difference (aFD) of maximum 1.4 frames (29 ms) for start of diastasis, down to 0.6 frames (12 ms) for mitral valve opening when performing a ten-fold cross-validation with test splits on triplane data from 240 patients. On an external independent test consisting of apical long-axis data from 180 other patients, the worst performing event detection had an aFD of 1.8~(30~ms). The proposed approach has the potential to significantly impact clinical practice by enabling more accurate, rapid and comprehensive event detection, leading to improved clinical measurements.
\end{abstract}


% \fi
\IEEEpeerreviewmaketitle

%% =====================================
%auto-ignore
%%===========================================
\section{Introduction}
\IEEEPARstart{C}{orrect} timing of the cardiac valve events is essential for many diagnostic measurements obtained by echocardiography. A robust and automated method would be beneficial, as it may enable more efficient workflows and minimize interobserver variability in the challenging task of determining the timing of valve events for measurements dependent on accurate cardiac timing. 

Most quantitative measurements of the left ventricle (LV) require the operator, or software in case of an automatic method, to set the timing of events in the cardiac cycle. Wall thicknesses and chamber dimensions are measured at end-diastole (ED) and end-systole (ES)~\cite{mitchell2019}, while functional measurements such as ejection fraction and myocardial strain use the difference between measurements at these timepoints~\cite{lang2015}. ED and ES are often defined depending on the modality and the measurements they are used for. ED is most commonly defined as the R-peak of the QRS-complex in a synchronized electrocardiogram (ECG) as this is usually easy to detect automatically, but this electrical signal tends to occur before LV filling has stopped fully. Volume-based measurements typically use timing based on maximum and minimum chamber volume, but they can also be defined by the timing of mitral and aortic valve closure (MVC, AVC), respectively~\cite{lang2015}. More comprehensive methods such as non-invasive myocardial work require not only timing of MVC and AVC, but also mitral and aortic valve openings (MVO, AVO)~\cite{russell2012} for which there are no established guidelines for timing measurements in 2D echocardiography. Although not currently in widespread clinical use, quantitative assessment of ventricular relaxation (diastolic function) could be facilitated by distinguishing the different phases of diastole, such as the rapid inflow, diastasis and atrial systole. Also, for left atrial strain measurements it is important to make a distinction between active and passive contraction, marked by atrial systole start (ASS)~\cite{badano2018}. \IEEEpubidadjcol{}

Conducting manual valve event timing measurements directly from 2D apical images is subjective and challenging, often due to sub-optimal image quality, but also other limitations such as the aortic valve not being visible in the apical four- and two-chamber (4CH and 2CH) views. According to previous studies, manual valve event timing can have a  significant interobserver variability~\cite{zolgharni2017,lane2021a}, with some studies presenting up to six frames average frame difference on definition of ED~\cite{lane2021a}. Valve event timings are commonly measured using pulsed wave (PW) Doppler spectrums, which show blood flow through the valves over time. The timing from these PW recordings are then applied to 2D recordings, using the electrocardiogram (ECG) signal to synchronize cycle start. This requires the user to both acquire additional recordings and perform the timing measurements, and introduces additional sources of error in the form of beat-to-beat variation, inconsistent ECG QRS-detection, and potential delays in PW Doppler signal relative to ECG~\cite{walker2019}. Despite these limitations, PW Doppler is favored for some measurements due to its better reproducibility~\cite{olsen2022a}. This highlights the importance of establishing high quality reference timing, as scoring well relative to interobserver in the most common clinical settings, is not sufficient for all clinical needs.

An automated method to detect the different cardiac phases through valve event timings directly from the 2D recordings could be beneficial by 1) eliminating all the issues with matching measurements from different recordings from different modalities 2) reducing the interobserver variability of measurements, and 3) enabling more efficient clinical workflows.

%%---------------------------------
\subsection{Prior work}
Numerous strategies have been proposed for automated detection of cardiac events utilizing ultrasound data. Kachenoura \textit{et al.}~\cite{kachenoura2007} proposed a method for ES detection, which involved the application of mitral valve motion and the correlation coefficient between the ED frame and any other frame within the cardiac cycle. Barcaro \textit{et al.}~\cite{barcaro2008} found the ED and ES frames from the volume extremities of a Simpson's Biplane approach enhanced with automatic segmentation. Furthermore, Aase \textit{et al.}~\cite{aase2011} introduced a method that employed speckle tracking to monitor specific points adjacent to the mitral annulus, and used the observed displacements to estimate the events and the duration of the cardiac cycle.

Deep learning approaches have also been applied to the specific task of detecting ED and ES from 2D recordings~\cite{dezaki2018,lane2021a,fiorito2018a}. These combine a convolutional neural network (CNN) to detect spatial features, with a temporal layer to get the changes over time. Lane \textit{et al.}~\cite{lane2021a} posed this as a regression problem, and used a pretrained ResNet50 to extract spatial features, followed by two layers of long-short term memory (LSTM) modules to capture the temporal features. Dezaki \textit{et al.} used a DenseNet backbone followed by two layers of gated recurrent units (GRU). In addition, they used a specialized loss based on the LV volume curve during the cardiac cycle, with ED and ES defined as the frames with maximum and minimum volume respectively~\cite{dezaki2018}. Fiorito \textit{et al.} incorporated temporal information into the initial parts of the network by utilizing 3D convolution layers followed by two layers of LSTM. They also modeled the prediction as a classification task, with events implied at the transitions between each phase~\cite{fiorito2018a}. Recently Reynaud \textit{et al.} used a transformer network to further enhance the spatio-temporal information extracted from the image sequences. However, the focus of the study was to perform automated ejection fraction calculation, hence event timing had a limited emphasis~\cite{reynaud2021}. To our knowledge, no prior studies have included automatic detection of the valve openings nor subphasing of diastole.

One notable limitation in existing research is that studies by~\cite{lane2021a,dezaki2018,reynaud2021} exclusively trained on the apical 4CH view. Additionally, \cite{fiorito2018a} also included 2CH view, but none of the studies included the apical long axis (APLAX) view, the only view with direct visibility of the aortic valve. Furthermore, only two events were taken into consideration in these studies. 

%%---------------------------------
\subsection{Main contributions}
We propose a novel method for automated detection of cardiac event timings directly from 2D apical images. To ensure accurate reference labels for training, we used apical triplane recordings for annotation. The triplane is a modality that concurrently records three 2D image planes, each oriented at a distinct angle relative to the others. The use of triplane data allowed us to get near-synchronized recordings of the different views from the same cycle, which is important when aortic valve events are only visible in the apical long axis (APLAX) view. The main contributions of this paper are
\begin{itemize}
    \item An improved approach for annotating valve events in echocardiography based on triplane recordings. 
    \item A training setup which uses synchronous apical images from multiple views to create a deep learning model which takes conventional apical images as input. 
    \item An automated timing method for all three standard apical views.
    \item The first automated approach for detecting six distinct cardiac events, namely the four different left-sided valve events, and the start of diastasis and atrial systole.   
    \item Systematic comparison to state-of-the-art methods, interobserver variability, as well as view specific observations. 
\end{itemize}

%auto-ignore
%%===========================================
\section{Methodology}
Previous publications have proven the viability of deep learning architectures for automated ED and ES detection~\cite{fiorito2018a,lane2021a,dezaki2018,reynaud2021}. We explored two such architectures that performed well on the two-event task, and expanded those to detect six distinct events on three different apical views. Both use a combination of convolutional layers to detect spatial image features followed by long short-term memory (LSTM) layers to account for the temporal changes from image to image.

%%----------------------------------
\subsection{Triplane annotation}
\begin{figure*}[tb]
    \centering
    \includegraphics[width=\textwidth]{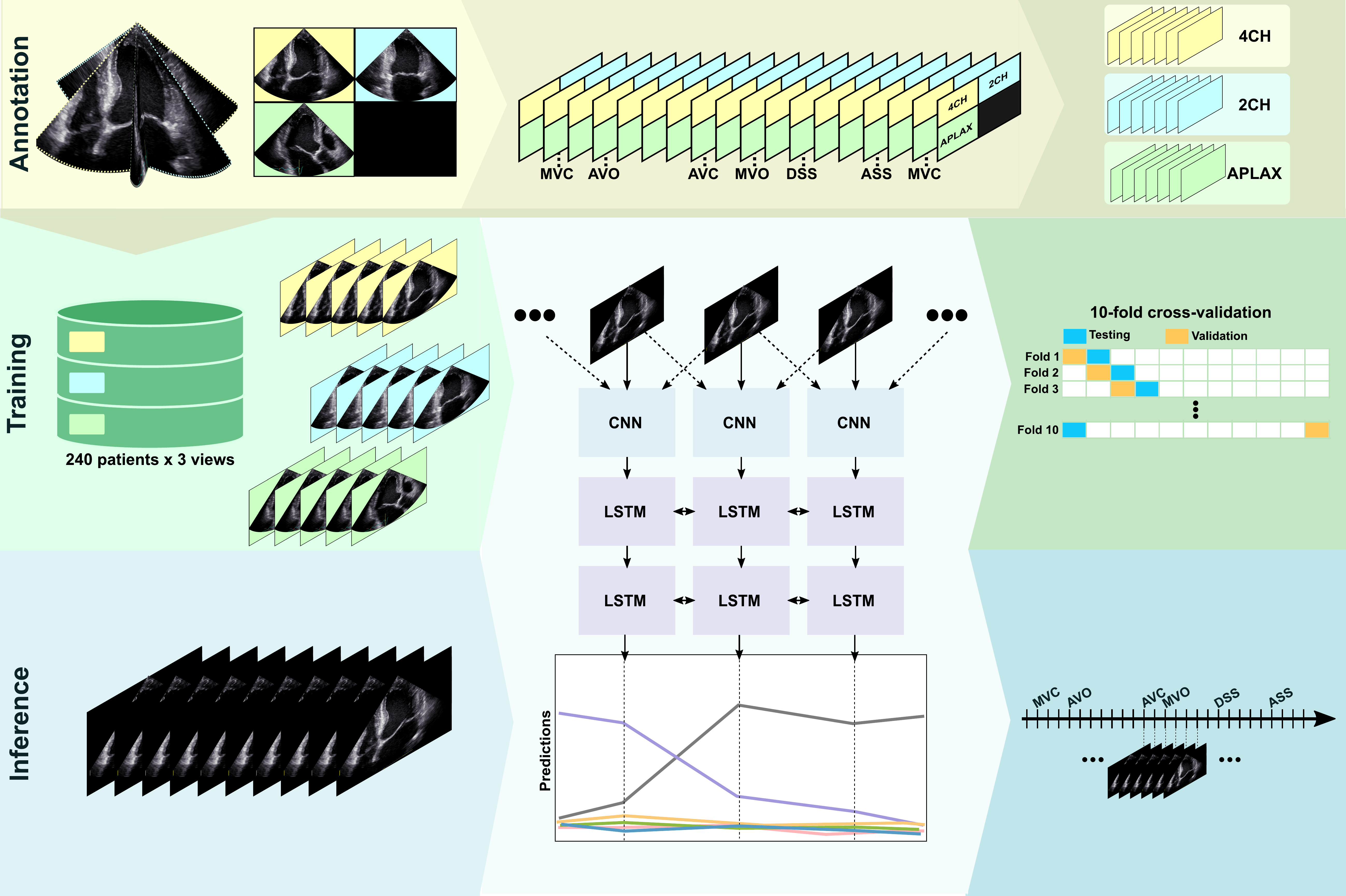}
    \caption{Annotations were performed on triplane data to provide the annotators with as much information as possible. The triplane views were then split and treated as three separate recordings with the same annotations before being used for training. The networks only train and evaluate on a single view at a time, so the resulting model is capable of predicting events from any regular single view 2D recording from any of the apical views.}
    \label{fig:annotation_pipeline}
\end{figure*}

Triplane recordings use a 3D probe to acquire data from three different image planes at the same time. In echocardiography, the most commonly used triplane mode is apical triplane, where the three main apical views, 4CH, 2CH and APLAX are recorded simultaneously for the same cardiac cycle. Annotation was performed with all three image planes visible, the triplane recordings were then split into the individual views with the same annotations before they were used for training or testing in the deep learning networks, as shown in \cref{fig:annotation_pipeline}. All triplane recordings were annotated by two independent experts to evaluate interobserver variability, to measure the uncertainty of the reference values.

Opening- and closure of the valves were simultaneously assessed in all three views. Mitral valve closure was defined as the frame where coaptation of the valve leaflets was visible in all three views. Opening was defined as the frame where coaptation was lost between the leaflets in at least one view. Each event was defined as follows:\\[-0.3cm]

\noindent \textit{Annotation protocol}\label{sec:annotation_protocol}
\begin{itemize}[labelindent=35pt,itemindent=0pt,leftmargin=*]
    \item[MVC:] First frame where the mitral valve was closed
    \item[AVO:] First frame where the aortic valve was open
    \item[AVC:] First frame where the aortic valve was closed
    \item[MVO:] First frame where the mitral valve was open
    \item[DSS:] First frame where the anterior mitral leaflet stopped its movement towards the left atrium after opening
    \item[ASS:] First frame where the anterior mitral leaflet moved towards the LV after diastasis
\end{itemize}
~\\[-0.3cm]
\noindent Definitions of MVC and AVC followed guideline recommendations~\cite{lang2015}. 

%%---------------------------------
\subsection{Classification network}
The classification network employed in this study was based on a modified version of the network architecture introduced by Fiorito \textit{et al.}~\cite{fiorito2018a}. This architecture incorporated 3D convolutional layers to incorporate temporal information into the spatial feature extraction. The method was extended to include six timing events. It consisted of five blocks of 3D convolutions layers, followed by batch normalization, ReLU activation and max pooling. The kernel size of the convolution layers was always 3 frames in the temporal axis, and (7 \texttimes\ 7) along the spatial axes of the first layer, followed by kernels of (3 \texttimes\ 3) in the following layers. The kernel size was kept equal to 3 for the temporal dimension. This resulted in a receptive field of 11 \texttimes\ 67 \texttimes\ 67 for the convolutional part of the network. The temporal size of 11 would effectively be further expanded by the LSTM layers, while the spatial size of 67 \texttimes\ 67 corresponds to a quarter of the total image area, providing coverage for the critical regions within the  frame, such as valves, annulus and base segments of the LV.

The number of filters was doubled for every convolutional layer, starting from 16. Pooling was only performed along the spatial axes, with a size of (2 \texttimes\ 2). The output of the CNN blocks was flattened and input into two subsequent LSTM layers with 32 cell units each. The output of the LSTM layers was input into a 1D convolution layer with a kernel size of 3. 

%%---------------------------------
\subsection{Regression Network}
The regression network was based on the work by Lane \textit{et al.}~\cite{lane2021a}. It used a ResNet50~\cite{he2015} convolutional network pre-trained on the ImageNet dataset. The CNN was followed by a LSTM-layer of size 2048, with a densely connected and ReLU activated layer reducing output size to 512 before the second LSTM-layer. The output of the LSTM was then flattened, and another dense network reduced output size to the final 6 labels with sigmoid activation applied to the output. Each of the 6 labels represented a valve event.

\begin{figure*}[tb]
    \centering
    \includegraphics[width=\textwidth]{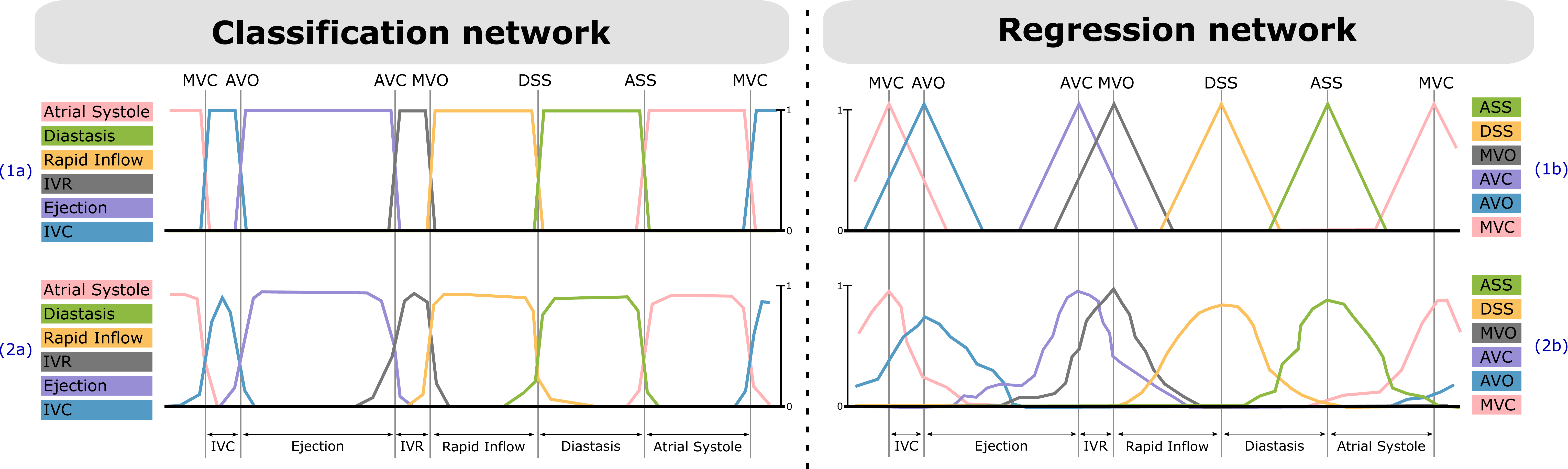}
    \caption{(1) Expert reference and (2) predicted output for the deep learning architectures. (1a) For the classification network, each frame is labeled by its phase. Each event timing represents the first frame of the corresponding phase. (2a) The events are derived from the phase predictions by finding the transitions from one phase to the next. (1b) The regression network labels each frame by its event, or its proximity to a nearby event. Each label is independent and weights are reduced linearly with a maximum distance of five frames. (2b) The events are derived from the predictions by finding the peak of each event label.}
    \label{fig:prediction_pipeline}
\end{figure*}

The original two-event network used a binary label linearly interpolated between ED and ES. However, this was unsuitable for accommodating additional events. Instead, one label was assigned to each event, and neighboring frames were labeled with the same label, albeit with lower weights further from the event. Our training used pseudo-labels that were linearly reduced from 1 at the event, to 0 at a 5 frame distance from the event, as illustrated in \cref{fig:prediction_pipeline}.

For inference, only a single pass of the video was used, with Gaussian smoothing of the prediction curves before detecting the events based on the tallest peaks.\\[-0.2cm]

The count of learnable parameters and floating point operations of the two networks are outlined in \cref{tab:network_complexity}.

\begin{table}[htb]
    \centering
    \caption{Network complexity metrics}
    \label{tab:network_complexity}
    \begin{tabular*}{\linewidth}{@{\hspace{1cm}}@{\extracolsep{\fill}}lll@{\hspace{1cm}}}
         \toprule
         & Classification & Regression \\
         \midrule
         Parameters & 1.7M & 60.3M \\
         FLOPs & 265M & 1021M\\
         \bottomrule
    \end{tabular*}
    \vskip 1mm
    {\raggedright \scriptsize FLOPs=Floating Point Operations \par}
\end{table}

%auto-ignore
\section{Experiments}
\subsection{Datasets}
Two different datasets were used to develop the methods: a triplane dataset used for training, validation and testing, and a separate external APLAX dataset used exlusively to test the trained model. An overview of the datasets used can be seen in \cref{tab:dataset_info}. Legal and ethical consent to use the data for research purposes was provided for all data used, detailed description of data and their ethical approval can be found at \cite{fermann2023}. % NB: manually edited bib-entry

\begin{table}[ht]
    \centering
    \caption{Overview of the different datasets}
    \label{tab:dataset_info}
    \begin{tabular*}{\linewidth}{l@{\extracolsep{\fill}}cc}
        \toprule
        & Triplane & External\\
        \midrule
        Patients & 240 & 180\\
        Avg. FPS & 48 & 61\\
        Views & 4CH, 2CH, APLAX & APLAX\\[0.1cm]
        Events & \parbox[c]{2.1cm}{MVC, AVO, AVC,\\ MVO, DSS, ASS} &  MVC, AVO, AVC, MVO\\[0.2cm]
        Use & Training/Testing & Testing\\
        \bottomrule
    \end{tabular*}
    \vskip 1mm%
    {\raggedright \scriptsize 4CH=Apical Four-Chamber, 2CH=Apical Two-Chamber, APLAX=Apical Long Axis, MVC=Mitral Valve Closure, AVO=Aortic Valve Opening, AVC=Aortic Valve Closure, MVO=Mitral Valve Opening, DSS=Diastasis Start, ASS=Aortic Systole Start, FPS=Frames Per Second \par}
\end{table}

\subsubsection{Internal triplane dataset}
In this study a total of 240 triplane recordings were utilized. The data were acquired at several different sites and fully anonymized before being annotated by two independent readers. The triplane data had an average sampling rate of 48 triplanes per second, and were acquired by alternating between each plane for a total of 144 image planes per second, using GE Vivid E95 ultrasound scanners equipped with 4Vc probes (GE Vingmed Ultrasound AS, Horten, Norway).

All valve events were annotated by visual assessment of the echocardiograms by two clinical experts, using EchoPAC release 204 (GE Vingmed Ultrasound AS, Horten, Norway). 

%%---------------------------------
\subsection{External APLAX dataset}
In addition to the triplane recordings, a secondary dataset was used for testing. This data consisted of 180 recordings with an average frame rate of 61 FPS and was acquired and annotated at a different site, Oslo University Hospital (OUS), by different expert observers. The external dataset consisted of 2D recordings from the APLAX view only. Although the 4CH view is more commonly used clinically and in prior work, the APLAX view was chosen because it has visibility of both the mitral and aortic valve. The DSS and ASS events were not included, but annotations followed the same protocol as described in \cref{sec:annotation_protocol} for the other four valve events.

%%---------------------------------
\subsection{Implementation details}
The modelling and experimentation were conducted on a workstation with an Ubuntu 20.04 operating system. The hardware consisted of an AMD Ryzen ThreadRipper 3970X with 32 cores and a clock speed of 3.70 GHz, 196 GB RAM and two NVIDIA RTX 3090 GPUs with 24 GB of memory each. Our choice of software included Tensorflow 2.11 for machine learning and Python 3.8 as the primary programming language.

%%---------------------------------
\subsection{Training and evaluation}
For training the classification network used a batch size of 8, with sequences randomly drawn from the data independent of views. All sequences for a given batch was padded to the length of the longest sequence of the individual batch. Thus, batches had varying sequence lengths and views. The image sequences were padded with zeros, while the labels were padded with -1. An Adam optimizer with a learning rate of 0.001, and a masked categorical cross entropy loss was used for training. The masking was conducted on labels with value equal to -1. The image shape of the inputs was (128 \texttimes\ 128) per frame, with 3 frames used as input for the CNN block. 

The regression network was trained with a batch size of 4, with sequences of fixed length 30 with a randomly selected starting point. Adam optimizer with a learning rate of 0.001, and a masked mean square error loss was used for training. The network was trained using bidirectional LSTMs and sigmoid activation. Image shape of the inputs was (112 \texttimes\ 112 \texttimes\ 3) to align with the pretrained ResNet-50 network.

Choice of sequence length was based on what performed best for each method, batch size differed due to memory limitations.
The models were trained using a ten-fold cross-validation with test splits on the triplane data with each split using 80\% for training, 10\% for validation and 10\% for testing. The different views for each recording were always kept in the same group to avoid data leakage. The maximum number of epochs for one cross-validation was 200, and early stopping with patience of 15 epoch was used. The results for the triplane dataset are based on the combined testing results for all folds, with models selected from validation scores for each fold, while the external APLAX results use an ensambled average of the 10 models. 

\subsection{Ablation experiments}
In order to assess the efficacy of the classification network, we conducted four ablation experiments. In the context of network design, we explored the impact of removing bidirectionality from the recurrent layers and substituting LSTM units with gated recurrent units (GRU). The subsequent two experiments addressed data manipulation. In one instance, we reduced the number of events from six to two (ED and ES), while the other experiment focused on training with a single acoustic view (4CH). Similar to the main setup, all ablation experiments were performed using the same protocol with cross-validation. 
%auto-ignore
%%===========================================
\section{Results}
Table~\ref{tab:interobserver} shows the interobserver variability in our triplane dataset together with results reported in a different study~\cite{lane2021a}. The average offset between R-peak of the ECG signal and annotated MVC was (\num{2.50\pm1.70}) ultrasound images.

% Inter-observer variability
\begin{table}[ht]
    \centering
    \caption{Interobserver variability comparing our triplane data with previously reported results from 4CH data~\cite{lane2021a}, measured in average frame difference (FD) with standard deviation, and average absolute frame difference (aFD)}
    \label{tab:interobserver}
    \begin{tabular*}{\linewidth}{l@{\extracolsep{\fill}}SSSS}
        \toprule
        & \multicolumn{2}{c}{Triplane (ours)} & \multicolumn{2}{c}{4CH (Lane \textit{et al.}~\cite{lane2021a})} \\
        \cmidrule{2-3} \cmidrule{4-5}
        {} & {FD} &    {aFD} & {FD} &    {aFD}\\
        \midrule
        MVC (ED)    &  0.02 (0.80) &    0.39&  -1.35 (1.31) &    1.55\\
        AVO         & -0.16 (0.71) &    0.34& ~ & ~\\
        AVC (ES)    &  0.29 (1.14) &    0.55& -0.90 (1.80) &    1.44\\
        MVO         & -0.24 (0.83) &    0.49& ~ & ~\\
        DSS         &  0.37 (1.79) &    1.01& ~ & ~\\
        ASS         & -0.19 (1.00) &    0.69& ~ & ~\\
        \bottomrule
    \end{tabular*}
    \vskip 1mm
    {\raggedright \scriptsize FD=Frame Difference, aFD=Absolute Frame Difference, ED=End-Diastole, ES=End-Systole, APLAX=Apical Long Axis, MVC=Mitral Valve Closure, AVO=Aortic Valve Opening, AVC=Aortic Valve Closure, MVO=Mitral Valve Opening, DSS=Diastasis start, ASS=Atrial Systole Start \par}
\end{table}

A subset of 30 recordings from the external dataset was also blindly annotated by one of the triplane annotators, with an average difference of 0.21 frames for the four valve events. This was higher than the interobserver difference between the two triplane annotators of 0.03 frames for the same events, but still indicates good annotation agreement. 

\Cref{tab:prev_methods} provides an overview of results for ED and ES detection, obtained from the two methods investigated in this study using the triplane and external datasets, alongside results presented in relevant literature for their respective datasets. External testing results were only available for one of the compared studies. A summary of average performance of the classification and regression networks, encompassing all events regardless of view, can be found in \cref{tab:classification_vs_regression}. 

\begin{table}[ht]
    \centering
    \caption{Average absolute frame difference of our 6-event models trained on triplane data compared to previously reported results for end-diastole (ED) and end-systole (ES). Our 3D CNN + 2-LSTM is refered to as the classification network, and ResNet-50 + 2-LSTM is refered to as the regression network. The datasets used in each study are different, both for training and external testing. In our study "training/testing" set refers to the triplane dataset used for training and testing with cross-validation, and "external testing" refers to the APLAX dataset which was collected and annotated at a different site by different experts.}
    \label{tab:prev_methods}
    \begin{tabular*}{\linewidth}{@{\extracolsep{\fill}\hspace{\tabcolsep}}lcccccc@{\hspace{\tabcolsep}}}
        \toprule
        & & \multicolumn{2}{c}{\hspace{-0.6em}Training/testing} & & \multicolumn{2}{c}{\hspace{-0.6em}External testing} \\
        \cmidrule(r){3-4} \cmidrule(r){6-7}
        Method & & {ED} & {ES} & & {ED} & {ES} \\
        \midrule
        3D CNN + 2-LSTM (ours)                  & & 0.74 & 0.93 & & 1.48 & 1.44\\
        ResNet-50 + 2-LSTM (ours)               & & 0.78 & 1.27 & & 1.99 & 1.62\\
        \midrule
        ResNet-50 + 2-LSTM \cite{lane2021a}     & & 0.66 & 0.81 & & 2.62 & 1.86\\
        3D CNN + 2-LSTM \cite{fiorito2018a}     & & 1.63 & 1.71 & & N/A & N/A\\
        Dense-Net + 2-GRU \cite{dezaki2018}     & & 0.20 & 1.43 & & N/A & N/A\\
        Video transformer \cite{reynaud2021}    & & 7.17 & 3.35 & & N/A & N/A\\
        \bottomrule
    \end{tabular*}
    \vskip 1mm
    {\raggedright \scriptsize CNN=Convolutional Neural Network, LSTM=Long Short-Term Memory, GRU=Gated Recurrent Unit, ED=End-Diastole, ES=End-Systole, APLAX=Apical Long Axis, MVC=Mitral Valve Closure, AVO=Aortic Valve Opening, AVC=Aortic Valve Closure, MVO=Mitral Valve Opening, DSS=Diastasis start, ASS=Atrial Systole Start, N/A=Not Available\par}
\end{table}

\begin{table}[ht]
    \centering
    \caption{Average prediction error for classification and regression method on the triplane dataset and external APLAX dataset. The values for each event are averaged over views for the triplane data. Measured in average absolute frame difference.}
    \label{tab:classification_vs_regression}
    \begin{tabular*}{\linewidth}{l@{\extracolsep{\fill}}cccc}
        \toprule
         & \multicolumn{2}{c}{\hspace{-1em}Classification} & \multicolumn{2}{c}{Regression} \\
         \cmidrule(r){2-3} \cmidrule(lr){4-5}
         & Triplane & External & Triplane & External\\
         \midrule
        MVC (ED)    & 0.74 & 1.48 & 0.78 & 1.99\\
        AVO         & 0.76 & 1.19 & 1.06 & 1.48\\
        AVC (ES)    & 0.93 & 1.44 & 1.27 & 1.62\\
        MVO         & 0.58 & 1.80 & 1.06 & 2.37\\
        DSS         & 1.39 & ~ & 2.29 & ~\\
        ASS         & 1.00 & ~ & 1.11 & ~\\
        \bottomrule
    \end{tabular*}
    \vskip 1mm
    {\raggedright \scriptsize ED=End-Diastole, ES=End-Systole, APLAX=Apical Long Axis, MVC=Mitral Valve Closure, AVO=Aortic Valve Opening, AVC=Aortic Valve Closure, MVO=Mitral Valve Opening, DSS=Diastasis start, ASS=Atrial Systole Start \par}
\end{table}

\Cref{tab:classification_vs_regression} shows that the classification network performs better than the regression method with an average 0.36 frames lower average absolute frame difference.
A detailed analysis of the classification network's performance for different events and views can be found in \cref{tab:classification_per_view} and \ref{tab:classification_per_view_ms}. \cref{tab:classification_per_view} presents the prediction error as average frame difference with standard deviation relative to manual annotations, while \cref{tab:classification_per_view_ms} displays the corresponding absolute frame differences in milliseconds.
\Cref{tab:ablation} shows the results from the ablation experiments comparing our method to variations in data and model design.

\begin{table}[ht]
    \centering
    \caption{Prediction method for each event and each view using the classification network measured in average frame difference with standard deviation, relative to expert reference timing.}
    \label{tab:classification_per_view}
    \begin{tabular*}{\linewidth}{l@{\extracolsep{\fill}}SSS}
        \toprule
        {} &  \multicolumn{1}{c}{4CH} &  \multicolumn{1}{c}{2CH} & \multicolumn{1}{c}{APLAX} \\
        \midrule
        MVC & -0.09 (1.25) &  -0.15 (1.29) &  -0.11 (1.00)\\
        AVO & -0.09 (1.27) &  -0.18 (1.36) &  -0.15 (1.27) \\
        AVC & 0.10 (1.49) &  0.17 (1.47) &  0.15 (1.39)\\
        MVO & -0.02 (1.12) &  0.04 (1.06) &  0.03 (1.10) \\
        DSS & -0.43 (2.70) &  -0.40 (2.45) &  -0.30 (2.12)\\
        ASS & -0.41 (2.98) &  -0.37 (2.30) &  -0.19 (1.27) \\
        \bottomrule
    \end{tabular*}
    \vskip 1mm
    {\raggedright \scriptsize 4CH=Apical Four-Chamber, 2CH=Apical Two-Chamber, APLAX=Apical Long Axis, ED=End-Diastole, ES=End-Systole, APLAX=Apical Long Axis, MVC=Mitral Valve Closure, AVO=Aortic Valve Opening, AVC=Aortic Valve Closure, MVO=Mitral Valve Opening, DSS=Diastasis start, ASS=Atrial Systole Start \par}
\end{table}

\begin{table}[ht!]
    \centering
    \caption{Prediction error for classification method for each event per view as average absolute frame difference in ms.}
    \label{tab:classification_per_view_ms}
    \begin{tabular*}{\linewidth}{l@{\extracolsep{\fill}}cccc}
        \toprule
        {} &  4CH &  2CH & APLAX & APLAX (External) \\
        \midrule
        MVC (ED) & 16 &  18 &  13 & 24\\
        AVO & 17 &  18 &  15 & 20\\
        AVC (ES) & 21 &  22 &  20 & 24\\
        MVO & 14 &  13 &  13 & 30\\
        DSS & 31 &  32 &  26 & ~\\
        ASS & 24 &  22 &  18 & ~ \\
        \bottomrule
    \end{tabular*}
    \vskip 1mm
    {\raggedright \scriptsize 4CH=Apical Four-Chamber, 2CH=Apical Two-Chamber, APLAX=Apical Long Axis, ED=End-Diastole, ES=End-Systole, APLAX=Apical Long Axis, MVC=Mitral Valve Closure, AVO=Aortic Valve Opening, AVC=Aortic Valve Closure, MVO=Mitral Valve Opening, DSS=Diastasis start, ASS=Atrial Systole Start \par}
\end{table}

\begin{figure}[htb]
    \centering
    \subfloat[Classification network]{%
        \includegraphics[width=\linewidth]{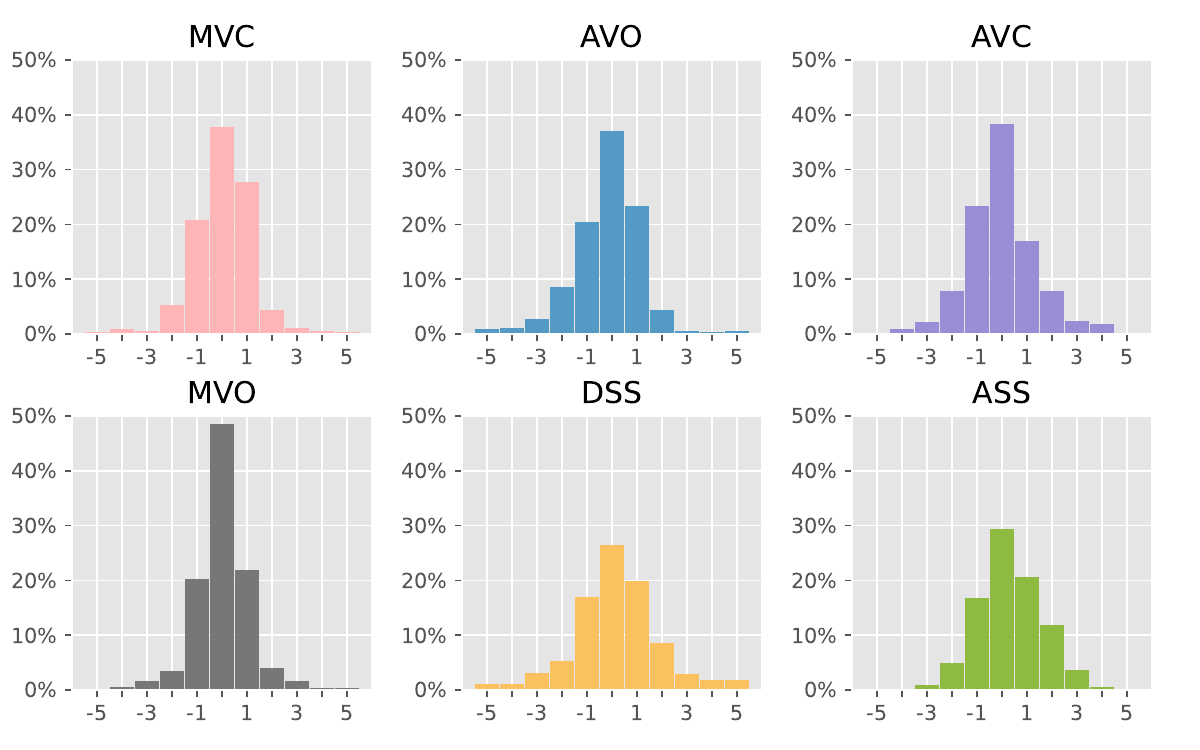}%
        \label{fig:hist:classification}%
    }
    \vspace{0.5cm}
    \subfloat[Regression network]{%
        \includegraphics[width=\linewidth]{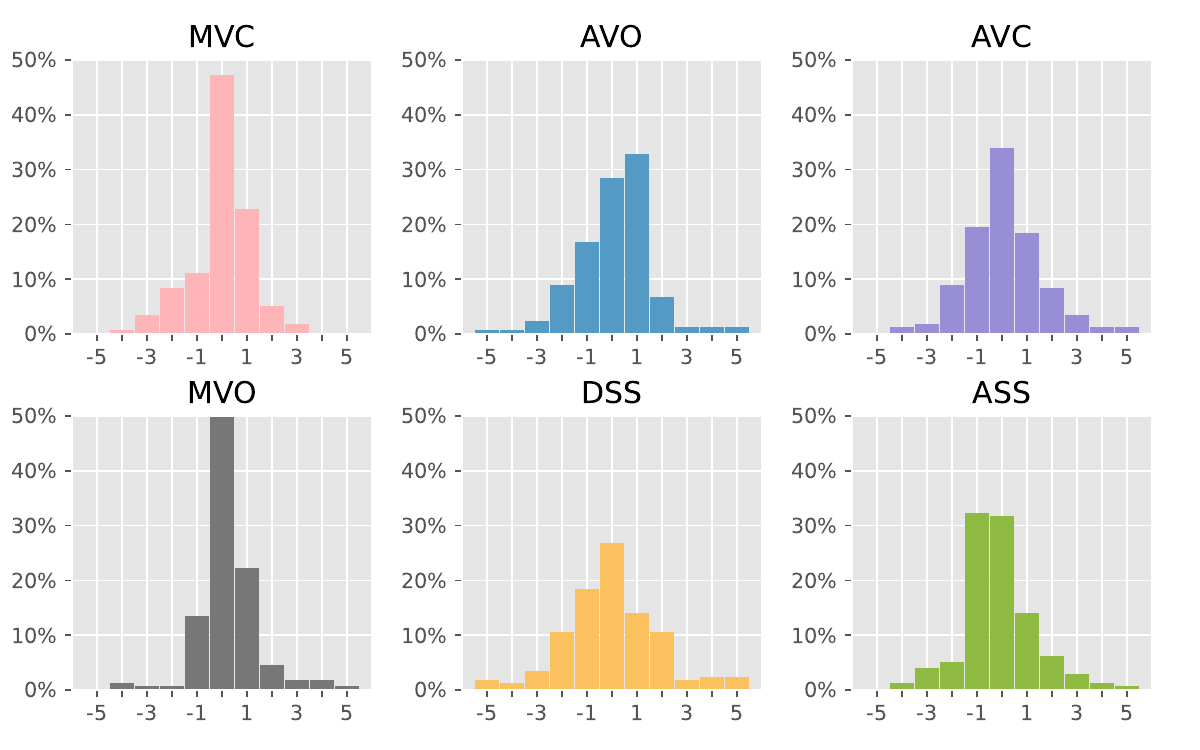}%
        \label{fig:hist:regression}%
    }
    \caption{Prediction error in frames for the \protect\subref{fig:hist:classification} classification network and \protect\subref{fig:hist:regression} regression network across all six different events. The histograms illustrate the proportional distributions of frame errors for each event.}
    \label{fig:hist}
\end{figure}

The spread of prediction errors for the different events can be seen in the histogram of \cref{fig:hist:classification} for the classification model and in \cref{fig:hist:regression} for the regression model. The histograms show that most of the results are reasonably normally distributed, with some skewed predictions for the regression method. 

\begin{table}[ht!]
    \centering
    \caption{Ablation study - Comparing results for classification network, prediction error in absolute frame difference}
    \label{tab:ablation}
    \begin{tabular*}{\linewidth}{l@{\extracolsep{\fill}}ccccc}
        \toprule
        ~ & Ours & Non-Bidir & Two-event & Single view & GRU\\
        \midrule
        MVC         & 0.74 & 0.82 & 0.80 & 0.92 & 0.74\\
        AVO         & 0.76 & 0.78 & ~ & 0.97 & 0.74\\
        AVC         & 0.93 & 1.01 & 1.52 & 1.67 & 0.93\\
        MVO         & 0.58 & 0.70 & ~ & 2.00 & 0.96\\
        DSS         & 1.39 & 2.62 & ~ & 3.31 & 2.66\\
        ASS         & 1.00 & 1.06 & ~ & 1.38 & 1.07\\
        \bottomrule
    \end{tabular*}
    \vskip 1mm
    {\raggedright \scriptsize GRU=Gated Recurrent Unit, MVC=Mitral Valve Closure, AVO=Aortic Valve Opening, AVC=Aortic Valve Closure, MVO=Mitral Valve Opening, DSS=Diastasis start, ASS=Atrial Systole Start \par}
\end{table}

The inference time for the automated methods per 60 frames is 82 ms for the classification network, and 147 ms for the regression network.

%auto-ignore
\section{Discussion}
This study showed that selecting triplane recordings for data annotation provides more context for the valvular motion compared to conventional 2D recordings. This can significantly mitigate the interobserver variability, thereby enhancing the confidence in the ground-truth data and promoting greater consistency in training labels. Furthermore we showed that it is possible to detect six distinct cardiac events from each cardiac cycle using 2D recordings of any apical view.

%%---------------------------------
\subsection{Triplane data and interobserver variability}
Absolute average interobserver variability across all events was 0.58 frames (12 ms), with highest values for the diastolic events DSS and ASS as shown in \cref{tab:interobserver}. At the start of diastasis, flow into the ventricle slows as the pressure equalizes and the valve movement stops, while atrial systole is marked by new movement after diastasis as the flow restarts. These are more subtle events to evaluate compared to the coaptation of leaflets used for valve openings and closures, which can explain why those particular events have higher variability.

Our interobserver variability for the ED and ES frames, respectively, was 0.39 and 0.55 in average absolute frame difference (aFD). This is significantly lower than what has previously been reported for event timing~\cite{zolgharni2017,lane2021a}, even when accounting the lower frame rate of triplane recordings. For example, Lane \textit{et al.} presents an average aFD of 1.55 and 1.44 for ED and ES detection, respectively, on apical 4CH recordings. Therefore, labeling triplane data and using the planes individually for training, allowing conventional use on 2D B-Mode images, is very attractive as the quality of annotations becomes higher than other approaches. Furthermore, the annotations are also performed within the same cardiac cycle. The default ED timing in most software solutions is also based on the R-peak of the ECG-signal whenever available, and in our triplane dataset it was found to be an average offset of (\num{2.50\pm1.70}) image frames between annotated MVC and the R-peak of the corresponding ECG. Mada \textit{et al.}~\cite{mada2015} found that changing the ED frame by \num{\pm 4} frames could change the global longitudinal strain by up to \qty{40}{\%}, highlighting the importance of selecting appropriate frames for each measurement.  

Another difference between the studies lies in the annotation protocol employed. For our study the emphasis was on the valves and their coaptation or movement, as described in \cref{sec:annotation_protocol}. Our focus when annotating was the timing itself, without considering if the frame would otherwise be suitable for 2D measurements. In contrast, the study conducted by Lane \textit{et al.}~\cite{lane2021a} involved the annotations for volume estimations, following the procedure used in preparation for Simpson's Biplane measurement in clinical practice, which constitutes a subtle difference from our definition. When ED and ES are defined by maximum and minimum volume respectively, they should generally be identical during the isovolumic phases, although ventricular deformation in practice leads to variations in the volume estimated from image cross-sections.

%%---------------------------------
\subsection{Detection of cardiac events}
Overall, the regression method was comparable to the classification method for most events, with slightly better accuracy for the classification method as shown in \cref{tab:classification_vs_regression}. The regression method also did not generalize as well when applied to the external APLAX data. These findings could indicate a mild overfitting of the regression method to the triplane data. Alternatively, it could imply that the spatio-temporal capacity of the 3D convolution-based network outperforms the regression setup.
To detect events, the classification network necessitates a transition from one phase to the next, whereas the regression network optimally utilizes the maxima of each prediction. This means the regression network has a more direct and intuitive relation between training labels and the physical events. On the other hand it necessitates more attention to avoid class imbalance and peak detection in noisy predictions. 

While the accuracy was lower on the external dataset, it also had a higher framerate than the triplane dataset. When comparing the average error in milliseconds, the difference was therefore reduced compared to error measured in frames. There are also some differences in image representation, where APLAX recordings occasionally have a tilted image sector and often a narrower, more focused field of view than what is found in the triplane data. And while the mitral valve is visible in APLAX, the annotation conditions are different with the lack of other views, which may impact the prediction results on the external dataset. A limitation of only using APLAX in the external data is that we do not have results with external data for 4CH and 2CH views, which could be different from our APLAX results. A random subset of the external data set was also annotated by one of the annotators for the triplane dataset and found no significant bias, which could otherwise have been a source of error.

The classification network had an average prediction error of 17 ms for the four valve opening and closure events. The DSS event which was most challenging to identify for the experts, also had the highest prediction error with an average of 30 ms across all apical views. 

As shown in \Cref{tab:classification_per_view} and \cref{tab:classification_per_view_ms}, it is possible to predict all cardiac events with comparable accuracy across the different views. Interestingly, despite the aortic valve events being characterized by valve changes that are not visible in the 4CH and 2CH views, the method is able to extract information necessary to achieve analogous results to those obtained using APLAX images. This aligns with previous findings, which demonstrated that the timing of AVC could be identified through the analysis of basal velocity traces derived from tissue Doppler images~\cite{aase2006}, or even through the motion of epicardium when measured with accelerometer~\cite{wajdan2022}.

\subsection{Network differences and ablation experiments}
In addition to resulting in more accurate results, the classification network was also lighter and faster than the regression network. Inference time was about 55 \% faster with the classification network and had fewer parameters to train as shown in \cref{tab:network_complexity}. This is in part explained by the use of a network pretrained on the ImageNet-database \cite{deng2009} which consists of a greater variety of color images. Training time was similar for the two networks. Recognizing the superior performance of the classification network, we prioritized ablation experiments specifically for it. 

The ablation experiments revealed that the use of bidirectional LSTM leads to superior results, aligning with our expectations. However, it is noteworthy that, excluding DSS, the non-bidirectional version demonstrated competitive performance for most events. Consequently, this alternative network configuration holds potential utility for prospective use.

Switching from LSTM to GRU layers produced comparable results, except for a deviation in the DSS event, where the performance degraded. The discrepancy in performance may stem from the inherent architectural differences between the two types of recurrent layers. LSTM networks have a more complex structure than GRU networks, and in the case of the DSS event class LSTM's ability to model long-term dependencies seem to be advantageous.

Transitioning from training with six events to two events in the dataset resulted in a degradation of performance, particularly in the detection of AVC. The absence of visibility of the aortic valve in the 4CH and 2CH views highlights the significance of having a dedicated class for MVO, which can aid in temporal contextualization. Conversely, training exclusively with the 4CH view led to a decline in results, likely attributed to the reduced dataset size and the inherent challenge of the aortic valve not being visible.

%%---------------------------------
\subsection{Comparison to prior work}
When comparing our findings to previously reported methods for ED and ES detection using machine learning approaches, as seen in \cref{tab:prev_methods}, we observe that our results are generally on par with the two-event methods. It is important to note the diversity in datasets across studies, and only one other investigation employed a test set that was distinct from the training set, rather than a subset of the same dataset.

Other studies predicted ED and ES timing mainly in the context of ejection fraction (EF) estimation using 4CH views, and annotations were made accordingly. While these definitions ideally align with events associated with valve motion, the divergent goal of EF estimation often results in different annotations in practice. For EF estimation, clinicians often prioritize frames with good image quality suitable for segmentation, and a minor deviation in time relative to valve closure may be acceptable for volume measurements. These factors, coupled with the lack of visible aortic valve in the 4CH view, pose challenges in direct comparisons to other datasets created for ED/ES detection.

Dezaki \textit{et al.}~\cite{dezaki2018} incorporated this relationship between valve events and ventricle volume in the training loss by using a non-linear regression loss based on volume curves. They have the largest relative difference in accuracy between ED and ES prediction, which is potentially due to the limitations of estimating ES from only the 4CH view. While modelling the timing loss of ED/ES based on volume curves is useful for EF estimations, it is not as generally applicable when we are adding more events that do not align with any volume extremes.  Lane \textit{et al.}~\cite{lane2021a} uses a linear regression loss and, during inference, enhanced accuracy by averaging several overlapping predictions. They also used a ResNet-50 pretrained on the ImageNet dataset, which includes a large amount of non-ultrasound images. The pretrained model served as a robust inital point for training, particularly valueable when thetraining data is limited. This approach had the best average accuracy for event prediction so it was a good candidate to expand to more events. Both \cite{dezaki2018} and \cite{lane2021a} model ED and ES as opposites in their loss, which is appropriate for the two-event task, but does not translate well to predicting six events. Although the architecture of our regression network is largely inspired by the work of Lane \textit{et al.}~\cite{lane2021a}, the event labeling, as shown in \cref{fig:prediction_pipeline} (1b), is distinctly different with each label representing only one event. There were some experiments with different shapes and widths of the soft event labels. The end result was a linear reduction in label weights for neighboring frames which is wide enough to avoid class imbalance, but steep enough to focus on the correct frame without too much noise and without averaging many overlapping predictions to get the event.

Reynaud \textit{et al.}~\cite{reynaud2021} used an initial neural network for dimensionality reduction, and then combined it with a transformer network that includes more spatio-temporal information at once. They reported prediction accuracy with average aFD of more than 7 frames for ED as shown in \cref{tab:prev_methods}. This was significantly less accurate than the other methods, so this architecture was not investigated further. However, this was also a study using timing as an intermediate step, with EF prediction as the main goal, and the network still performed well for that purpose.

The 3D CNN architecture presented by Fiorito \textit{et al.}~\cite{fiorito2018a} reported an average aFD of 1.63 and 1.71 for ED and ES respectively. Their method added temporal information into the initial convolutional network, but they also used a different labeling and classification method which seemed particularly well suited for our extension to additional events. Splitting the cardiac cycle into additional physiologically distinct phases means that there is a narrower range of states to fit into each class. For that reason Fiorito \textit{et al.}~\cite{fiorito2018a} was the inspiration for our classification network which diverged very little from this original architecture beyond the extension to multiple classes. The updated version, however, had significant improvement in reported accuracy from 1.63 to 0.78 aFD for ED from the original two-event version to our six-event network.

\subsection{Limitations}
Although our method reports good results on our dataset, comparisons to other studies are very limited due to the changes in annotated events and ultrasound views. The use of triplane data for annotation and training is novel, but future use and validation of this method is difficult without publicly available data sets. Future work using triplane data should strive to create a shareable data set.

Our strategy for valve event timing involves leveraging known machine learning methods for the two-event problem and expanding them to include more events, views and utilization of triplane data for annotation. The field of machine learning evolves fast, and there are many interesting solutions to adjacent problems  that warrant exploration in the future. For instance, there is a wide range of different pathologies that can affect the representation of the ventricles, valves and the cardiac cycle which can be difficult to capture beyond having a varied data set. Contrastive learning is a technique that has shown great promise in clustering similar data, with Xiao \textit{et al.}~\cite{xiao2023} designing a method to incorporate both high-level and low-level semantic information through deep contrastive learning applied to time-series data. Such an approach could potentially be used to accommodate a large variety of pathology and their implications on valve event timing. 

Our novel approach to annotation relies on triplane recordings, which are not as widely used as 2D recordings in echocardiography and limits the amount of data we were able to gather for our dataset. Generating synthetic triplane data based on recordings from other modalities could be a way of accessing larger training datasets while ensuring all the data is physiologically sound~\cite{tiago2023a}.

%%---------------------------------
\subsection{Limitations with respect to relevant pathologies}
Since training data was anonymized, we lack comprehensive clinical information about the study participants. Aortic sclerosis, and especially more severe stenosis, is a prevalent valve disease that could complicate correct assessment of AVO and AVC from B-mode images of severely stenotic aortic valves. Nevertheless, based on the subjective experiences of the annotators, the dataset did not contain a significant number of valvular stenosis cases.

Although ground truth for all events is defined by valve motion, both clinicans and deep learning methods also rely on ventricular motion patterns. Left bundle branch block (LBBB)-induced dyssynchrony is another challenge that has not been assessed. Without including known LBBB-cases in the training set, it is plausible that the timing accuracy for patients with dyssynchronous contraction patterns will be lower. The accuracy of predicting aortic valve events on apical 2CH and 4CH views suggests that ventricular motion, and not just valve movement, is important for generating accurate predictions.

Arrhythmia makes it challenging to apply Doppler-based event timing to separate 2D recordings due to the inconsistent cycle time/RR-interval. Especially for atrial fibrillation where cycle times can be highly inconsistent, these methods will likely not provide correct timing of the cardiac phases. In our dataset we have 28 subjects where it was not possible to annotate atrial systole. Although we did not have ECGs or clinical diagnosis of atrial fibrillation, it is likely that these subjects had atrial fibrillation during echocardiography. 

Due to lack of diagnostic information, we compared the timing intervals of our training data to normal ranges established by TDI~\cite{biering-sorensen2016}. Our findings revealed that the average was within normal ranges for isovolumic relaxation time (IVRT), with an average of 107 ms, the ejection time with an average of 287 ms. However, the isovolumic contraction time, with an average of 58 ms, exhibited a longer duration than the typical range. \Cref{tab:normal_intervals} shows the distribution of patients with different phase intervals, suggesting that the dataset contained mixed pathologies as we would expect. It should be noted that 2D recordings are not suitable for measuring the length of intervals for individual patients due to the low temporal resolution, but in aggregate it represents an estimate of the distribution.

\begin{table}[htb]
    \centering
    \caption{Proportion of patients with cardiac intervals within normal ranges as defined by Biering-Sørensen \textit{et al.}~\cite{biering-sorensen2016}}
    \label{tab:normal_intervals}
    \begin{tabular}{lccc}
        \toprule
         &  < Normal & Normal & > Normal\\
         \midrule
         IVRT (72-112 ms) & 19\% & 59\% & 21\% \\
         IVCT (23-49 ms) & 8\% & 39\% & 53\%\\
         ET (272-314 ms) & 21\% & 29\% & 50\%\\
         \bottomrule
    \end{tabular}
\end{table}

Futher, we calculated LV EF for the dataset using an algorithm for automated volume measurements based on deep learning-based segmentation~\cite{smistad2020}. This was conducted using the ED and ES events set by the annotators, and Biplane Simpson's from the apical 4CH and 2CH planes. From this, the EF was calculated to be (\qty{38\pm 13}{\%}) in our training dataset, indicating a mix of healthy and pathological cases. Hence, the results holds for a wide range of patient conditions.

%auto-ignore
\section{Conclusion}
In this paper we presented a novel approach for automatic timing of six distinct cardiac valve events in echocardiography using deep learning. We showed that annotations of triplane recordings exhibit low interobserver variability in comparison to alternative approaches, and contend that this is correlated with high quality training data. Results were on par with state-of-the-art, but also expand upon it by incorporating four additional cardiac events. We believe these types of fully automated approaches have the potential to enhance existing clinical measurements and to enable new ones through robust timing with low variability, and time savings through more efficient workflows.

%% =====================================

\section*{Acknowledgment}
The authors would like to thank Helge Skulstad at Oslo University Hospital for collaboration on data collection. We also extend our gratitude to Adrian Meidell Fiorito for his valuable methodological contributions during his master's thesis project.

\ifCLASSOPTIONcaptionsoff
  \newpage
\fi


% \bibliographystyle{IEEEtran}
% \bibliography{IEEEabrv,EventTiming_bibtex.bib}

\printbibliography

%   I'd rather skip this personally - BSF
%\vfill

%\enlargethispage{-15in}

\end{document}